\documentclass[conference,a4paper]{APSIPA2021}
\usepackage{amsmath,graphicx}
\usepackage{listings}
\usepackage{algorithm}
\usepackage{algpseudocode}
\usepackage{booktabs}
\usepackage{multirow}
\usepackage{siunitx}
\usepackage{url}
\usepackage{scalefnt}
\usepackage{amsfonts}
\usepackage{cite}
\usepackage{array}
\usepackage{longtable}
\usepackage{colortab}
\usepackage{colortbl}
\usepackage{arydshln}

\usepackage{graphicx}
\graphicspath{ {./images/} } 

\def\L{{\cal L}}

\makeatletter
\def\adl@drawiv#1#2#3{%
        \hskip.5\tabcolsep
        \xleaders#3{#2.5\@tempdimb #1{1}#2.5\@tempdimb}%
                #2\z@ plus1fil minus1fil\relax
        \hskip.5\tabcolsep}
\newcommand{\cdashlinelr}[1]{%
  \noalign{\vskip\aboverulesep
           \global\let\@dashdrawstore\adl@draw
           \global\let\adl@draw\adl@drawiv}
  \cdashline{#1}
  \noalign{\global\let\adl@draw\@dashdrawstore
           \vskip\belowrulesep}}
\makeatother

\usepackage{geometry}
\usepackage{fancyhdr}

\title{ESPnet-ONNX: Bridging a Gap Between\\Research and Production}

\author{%
\authorblockN{%
Masao Someki\authorrefmark{1},
Yosuke Higuchi\authorrefmark{2},
Tomoki Hayashi\authorrefmark{3}\authorrefmark{4},
Shinji Watanabe\authorrefmark{5}
}
\authorblockA{
\authorrefmark{1}
IBM Japan Ltd., Tokyo, Japan  Email: masao.someki@gmail.com\\
\authorrefmark{2}
Waseda University, Tokyo, Japan Email: higuchi@pcl.cs.waseda.ac.jp\\
\authorrefmark{3}
Nagoya University, 
\authorrefmark{4}
Human Dataware Lab Co., Ltd., Nagoya, Japan \\  Email: hayashi.tomoki@g.sp.m.is.nagoya-u.ac.jp\\
\authorrefmark{5}
Carnegie Mellon University, Pittsburgh, USA Email: shinjiw@ieee.org
}
}

\begin{document}
%
\maketitle
\begin{abstract}
In the field of deep learning, researchers often focus on inventing novel neural network models and improving benchmarks.
In contrast, application developers are interested in making models suitable for actual products, which involves optimizing a model for faster inference and adapting a model to various platforms (e.g., C++ and Python).
In this work, to fill the gap between the two,
we establish an effective procedure for optimizing a PyTorch-based research-oriented model for deployment,
taking ESPnet, a widely used toolkit for speech processing, as an instance.
We introduce different techniques to ESPnet, including converting a model into an ONNX format, fusing nodes in a graph, and quantizing parameters, which lead to approximately 1.3-2$\times$ speedup in various tasks (i.e., ASR, TTS, speech translation, and spoken language understanding) while keeping its performance without any additional training.
Our ESPnet-ONNX will be publicly available at \url{https://github.com/espnet/espnet_onnx}.
\end{abstract}
\section{Introduction}
\label{sec:intro}

Advances in deep learning have led to remarkable success in end-to-end speech processing,
including Automatic Speech Recognition (ASR), Text-to-Speech (TTS), Speech Translation (ST), and Spoken Language Understanding (SLU).
Transformer~\cite{vaswani2017attention} and Conformer~\cite{gulati2020conformer} have achieved promising results in various speech tasks~\cite{dong2018speech,karita2019comparative,radfar2020end,guo2021recent},
thanks to their sophisticated Deep Neural Network (DNN) architecture and large model capacity.
Self-supervised models~\cite{baevski2019vq,baevski2020wav2vec,hsu2021hubert}
leverage a vast amount of unlabeled audio data to learn versatile representations,
which have shown to boost the performance of downstream speech tasks~\cite{yang2021superb}.
However, these approaches tend to increase the number of model parameters,
making the inference process computationally intensive and, thus, unsuitable for actual products (e.g., run on edge devices).
As system latency is an essential factor in real-world deployments,
much effort has been made to speed up the inference of DNN models while maintaining task performance~\cite{kuchaiev2019nemo,yao2021wenet}.

There are several choices for improving the inference speed of speech processing models,
such as
parameter reduction,
non-autoregressive modeling, and
model optimization.
Parameter reduction can be realized as post-processing of model training.
For example,
parameter pruning explores the redundancy among neurons, and
removes unimportant neurons, layers, or attention heads that are less active for solving a target task~\cite{hanson1988comparing,lecun1989optimal,hassibi1992second,narang2017exploring,lee2021layer,voita2019analyzing,michel2019sixteen}.
Another is to train a smaller model from scratch,
using parameter sharing~\cite{li2019improving} or knowledge distillation~\cite{pang2018compression,takashima2018investigation}.
While these approaches have shown promising results,
reducing model capacity often leads to inevitable performance degradation.
In addition, they often require special treatments (e.g., architecture changes, retraining, and optimization tricks) during training.
Non-autoregressive models simultaneously generate multiple outputs in a sequence~\cite{graves2014towards,gu2017non},
which significantly speeds up the inference compared to autoregressive models.
It has been actively studied in ASR~\cite{chen2020non,chan2020imputer,higuchi2020mask,higuchi2021comparative} and TTS~\cite{kumar2019melgan,ren2020fastspeech,kim2021conditional},
enabling high-quality sequence generation while fastening the inference speed.
These approaches have also been shown effective in ST~\cite{inaguma2021orthros} and SLU~\cite{omachi2022non}.
However, the performance of non-autoregressive models is still limited to some extent
in that there are difficulties in combining with an external language model and beam-search decoding.

In contrast to the above methods that designed and trained a model itself,
fast inference can be achieved by optimizing a model to a specific environment (hardware) where it is run on.
For example, TensorRT~\cite{tensorrt} facilitates high-performance inference on NVIDIA GPUs
through optimizations, including layer/tensor fusion and memory caching.
Open Neural Network Exchange (ONNX)~\cite{bai2019} provides a flexible conversion between different deep learning frameworks,
which makes it easier to access hardware optimizations built for a specific framework.
ONNX also provides graph optimizations to further improve inference speed by graph-level modifications, such as node elimination and node fusion.
Since these optimization approaches keep the same number of parameters and model architecture,
one can safely speed up the inference while mitigating the performance drop in task performance.

In this work, we investigate the effect of conversion to the ONNX format for various speech tasks.
We also applied two optimization techniques, node fusion and quantization, to examine the effect of model optimization.
We used the model built with the End-to-End Speech Processing Toolkit (ESPnet)~\cite{watanabe2018espnet}, a widely used toolkit for speech processing.
This work does not require any additional training on the pretrained model,
so it can be applied to over 278 pretrained models of ESPnet publicly available at \url{https://huggingface.co/espnet}.
The key contributions and findings of this work are summarized as follows:
\begin{itemize}
    \item We demonstrate that conversion and optimizations provides approximately 2$\times$ speedup in various tasks, including ASR, TTS, ST, and SLU.
    \item We report that conversion can speed up fine-tuning process of pre-trained self-supervised model.
    \item We open-source our ESPnet-ONNX\footnote{Released at \url{https://github.com/espnet/espnet_onnx}},
    which enables users to convert over 278 models instantly.
\end{itemize}

\section{ONNX Conversion and Model Optimization}
\label{sec:conversion}

In this section, we describe our procedure for optimizing a speech processing model for deployment:
conversion of a PyTorch-based model to ONNX format and
model optimization using node fusion and quantization.

\subsection{ONNX Conversion}
\label{ssec:conversion}


\begin{figure}[t]
  \begin{minipage}[b]{0.45\linewidth}
    \centering
    \includegraphics[width=1.0\linewidth]{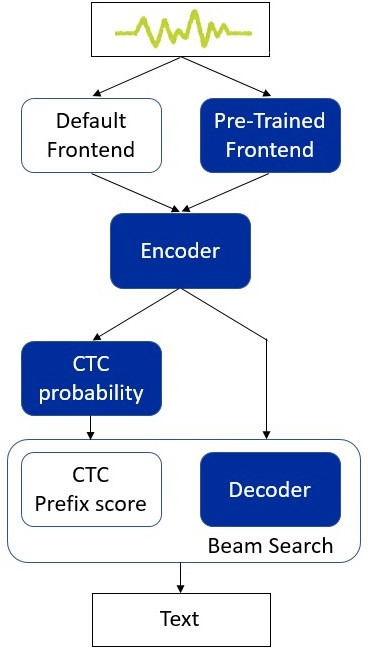}
    {\small (a) ASR}
  \end{minipage}
  \begin{minipage}[b]{0.45\linewidth}
    \centering
    \includegraphics[width=1.0\linewidth]{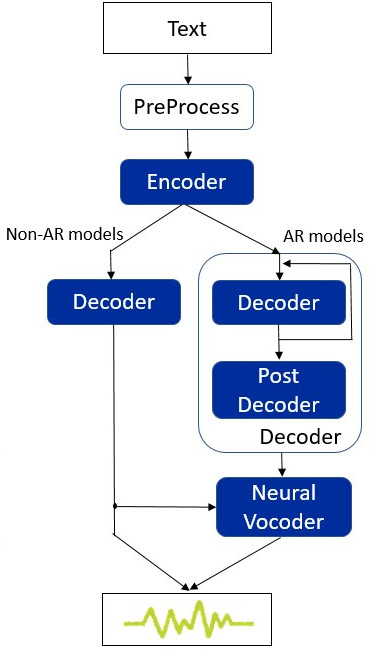}
    {\small (b) TTS}
  \end{minipage}
    \caption{Schematic diagrams of converting an ASR and TTS system. For TTS, we show autoregressive (AR) and non-AR decoders. The modules in blue can be converted to the ONNX format.}
  \label{fig:torch_and_onnx}
\end{figure}

\subsubsection{ASR}
\label{sssec:conversion_asr}
Since ONNX defines computational graphs statically before a model runs,
a PyTorch-based ASR model (Fig.~\ref{fig:torch_and_onnx}(a)) cannot easily be converted into an ONNX format.
The model contains dynamic loop computation (DLC) depending on the input length (e.g., autoregressive calculation).
For example, the beam search algorithm requires iterative CTC and decoder calculations.
Here,
the number of loops in beam search depends on the length of an input sample,
which is unknown during the conversion process.
This makes it difficult to convert the beam search process into a static graph.
The attention mechanism, on the other hand,
can treat a flexible input length, as it works only on a single forward pass.
To prevent the occurrence of such DLC in a static graph,
we divide the system into modules (boxes in Fig.~\ref{fig:torch_and_onnx}),
each of which is whether converted to ONNX or kept the same.
Note that the CTC calculation is separated into two modules because the computation of the prefix score contains DLC process.
Therefore, we only convert the computation of CTC probability into ONNX.

The above conversion process can also be used for ST and SLU models,
as they basically follow the same architecture as that of ASR.

\subsubsection{TTS}
\label{sssec:conversion_tts}
A TTS model can also be converted to the ONNX format by dividing a model into modules and avoiding DLC.
For example, in the case of an autoregressive model such as Tacotron2~\cite{shen2018natural},
as shown in Fig. \ref{fig:torch_and_onnx}(b),
the number of iterations for the decoder process depends on the input text length.
Therefore, we need to split the decoder into two modules to avoid the DLC process:
The autoregressive module and the post-decoder module that does not require iteration.
For a non-autoregressive model, since there is no DLC process as shown in Fig.~\ref{fig:torch_and_onnx}(b),
we can convert the whole model into a single ONNX model.
This way, the same procedure of ASR conversion can be applied to TTS.

\subsection{Node Fusion}
\label{ssec:node_fusion}

\begin{figure}[t]
  \begin{minipage}[b]{0.45\linewidth}
    \centering
    \includegraphics[width=1.0\linewidth]{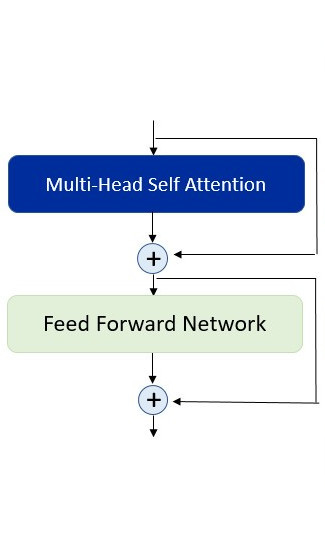}
    {\small (a) Transformer}
  \end{minipage}
  \begin{minipage}[b]{0.45\linewidth}
    \centering
    \includegraphics[width=1.0\linewidth]{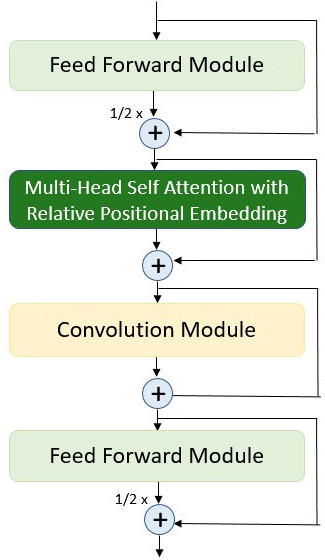}
    {\small (b) Conformer}
  \end{minipage}
  \caption{An overview of Transformer and Conformer. Node fusion is applied to the Attention, Attention with Relative Positional Embedding, and Layer Normalization.}
  \label{fig:trf_and_cfm}
\end{figure}

Node fusion is one of the techniques to optimize a graph;
for an ONNX model, it is provided by ONNX Runtime~\cite{onnxruntime}.
Transformer~\cite{vaswani2017attention} and Conformer~\cite{gulati2020conformer},
widely used architectures for speech processing tasks,
are composed of stacked identical blocks containing multiple modules (e.g., Attention and Layer Normalization).
Therefore, a specific shape appears many times in a graph of such models.
Node fusion combines this particular sub-graph into a single node and computes it efficiently using multi-threading techniques.

\begin{figure}[!t]
    \centering
    \includegraphics[width=0.8\linewidth]{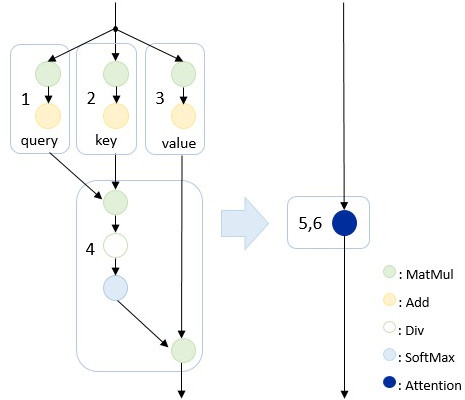}
    \caption{Simplified graph of Attention layer without reshape nodes (left) and its fused graph (right). Each number represents the equation number.}
    \label{fig:node_fusion}
\end{figure}

\subsubsection{Transformer}
\label{sssec:fusion:transformer}

Each Transformer can be divided into two components as shown in Fig. \ref{fig:trf_and_cfm}(a):
Multi-Head Self Attention and Feed Forward Module.
ESPnet uses Scaled Dot-Production Attention~\cite{vaswani2017attention} in the Transformer and 
converting this attention yields the sub-graph shown on the left side of Fig. \ref{fig:node_fusion}.
Note that the nodes related to tensor transformation and reshaping are ignored for simplicity.
Here, let the $L$-length input sequence for the attention module be $X \in \mathbb{R}^{L \times D}$ and 
the scale $d_k$ be $D / H$, where $D$ is the hidden dimension and $H$ is the number of heads.
The queries, keys, and values of the attention calculation are packed in matrices $Q$, $K$, and $V$, respectively.
The output of the attention for the original ONNX model (before node fusion) is calculated as follows:
\begin{align}
    Q &= \mathrm{MatMul}(X, W^Q) + \mathrm{bias}^Q, \label{eq:query} \\
    K &= \mathrm{MatMul}(X, W^K) + \mathrm{bias}^K, \label{eq:key}\\
    V &= \mathrm{MatMul}(X, W^V) + \mathrm{bias}^V, \label{eq:value} \\
    \mathrm{Output} &=
    \mathrm{Attention}(
        Q,
        K,
        V
    ) \nonumber \\
    &= \mathrm{Softmax}(\frac{QK^{\top}}{\sqrt{d_k}})V, \label{eq:attention}
\end{align}
where $W^Q \in \mathbb{R}^{D\times D}$, $W^K \in \mathbb{R}^{D\times D}$, $W^V \in \mathbb{R}^{D\times D}$, $\mathrm{bias}^Q \in \mathbb{R}^D$, $\mathrm{bias}^K \in \mathbb{R}^D$, $\mathrm{bias}^V \in \mathbb{R}^D$ represent the weights and biases of queries, keys, and values, respectively.

In the fused node shown on the right side of Fig. \ref{fig:node_fusion},
we compute the above equations in a multi-threading manner.
For $h = 0, \cdots, H - 1$, the output of the fused node is calculated as follows:
\begin{align}
    \mathrm{Output} &=
    \mathrm{Concat}(
    \mathrm{Attention}(
        Q_h,
        K_h,
        V_h
    ))\nonumber \\
    &= \mathrm{Concat}(\mathrm{Softmax}(\frac{Q_{h}K^{\top}_h}{\sqrt{d_k}})V_h), \label{eq:attention_h}
\end{align}
where $Q_h$, $K_h$, and $V_h$ for each thread is calculated as follows:
\begin{align}
    Q_h, K_h, V_h = \mathrm{Split}&(
        \mathrm{MatMul}(X_h, \mathrm{Concat}(W^Q_h, W^K_h, W^V_h)
    ) \nonumber \\
    &+ \mathrm{Concat}(\mathrm{bias}^Q_h, \mathrm{bias}^K_h, \mathrm{bias}^V_h)), \label{eq:qkv}
\end{align}
where $W^Q_h \in \mathbb{R}^{D\times d_k}$, $W^K \in \mathbb{R}^{D\times d_k}$, $W^V \in \mathbb{R}^{D\times d_k}$, $\mathrm{bias}^Q_h \in \mathbb{R}^{d_k}$, $\mathrm{bias}^K_h \in \mathbb{R}^{d_k}$, and $\mathrm{bias}^V_h \in \mathbb{R}^{d_k}$ are the weights and biases of queries, keys, and values for each thread, respectively.
As described in the two equations, Eqs. (\ref{eq:attention_h}) and (\ref{eq:qkv}),
the fused node calculates simultaneously in $H$ threads.
Therefore, the computation can be performed more efficiently compared to the original model if the device is capable of computing with multiple threads.

\subsubsection{Conformer}
\label{sssec:fusion:conformer}
Conformer is a variant of Transformer augmented with convolution,
as shown in Fig. \ref{fig:trf_and_cfm}(b).
The important change that has a significant influence to the node fusion is the replacement of the Attention component
with Attention with Relative Positional Embedding (RelPosAttention).
Relative Positional Embedding was originally proposed in Natural Language Processing~\cite{shaw-etal-2018-self},
and Conformer utilizes the embedding proposed in Transformer-XL~\cite{dai-etal-2019-transformer}.
In the Transformer-XL paper, they introduced a Relative Positional Embedding $R_{i-j}$,
which considers the relative distance between a query vector at position $i$ and a key vector at position $j$.

For speech processing tasks, since a model can see the future information, the range of $i-j$ is from $-(L-1)$ to $L-1$.
In the fused RelPosAttention node,
we need to calculate the following equations with $H$ threads:
\begin{align}
    \mathrm{rel}_{i-j} &= \mathrm{ac} + \mathrm{bd}, \label{eq:ac_plus_bd}\\
    \mathrm{ac} &= X^{\top}W^{Q\top}W^{K\top}X + u^{\top}W^{K\top}X, \nonumber \\
    \mathrm{bd} &= X^{\top}W^{Q\top}W^{R\top}R_{i-j} + v^{\top}W^{R\top}R_{i-j}, \nonumber
\end{align}
where $u \in \mathbb{R}^{H \times d_k}$ and $v \in \mathbb{R}^{H \times d_k}$ are the bias vectors,
$R \in \mathbb{R}^{L \times L}$ and $W^R$ are the given relative positional embedding and its projection weight,
and $\mathrm{ac} \in \mathbb{R}^{L \times L}$ and $\mathrm{bd} \in \mathbb{R}^{L \times L}$.
ESPnet uses the Relative Position Embedding $R^{En} \in \mathbb{R}^{2L-1 \times D}$ to represent $R_{i-j}$, 
where $R^{En}_{L-1}$ shows the position 0.
So we need a specific process to fit the size of the tensor $R^{En}$ to $\mathbb{R}^{L \times L}$
to calculate Eq. (\ref{eq:ac_plus_bd}).
ESPnet processes this tensor transformation by applying padding, reshaping, and slicing the tensor.
However, converting this process would generate many nodes related to shape modification,
which is not suitable for multi-threading.

Here, let the weight of $R^{En}_h \in \mathbb{R}^{2L-1 \times d_k}$ be $W^{R^{En}}_h \in \mathbb{R}^{d_k \times d_k}$, 
the $\mathrm{rel}_{i-j}$ for each thread $\mathrm{rel}_{h,i-j}$ is calculated as:
\begin{align}
    \mathrm{rel}_{h,i-j} &= \mathrm{ac}_h + \mathrm{bd}_{h,i-j}, \\
    \mathrm{bd}_h &=X^{\top}_hW^{Q\top}_hW^{R^{En}\top}_hR^{En}_h
    + v^{\top}_hW^{R^{En}\top}_hR^{En}_h, \\
    \mathrm{bd}_{h, i-j} &= 
    \begin{bmatrix}
        \mathrm{bd}_{h, L-1} &  \cdots & \mathrm{bd}_{h, 2L-1} \\
        \vdots  & \ddots  & \vdots \\
        \mathrm{bd}_{h, 0} & \cdots &  \mathrm{bd}_{h, L-1} \\
    \end{bmatrix}.
\end{align}
By calculating the tensor transformation process in this way,
most of the nodes can be fused
and computed efficiently with the multi-threading technique.


\subsection{Quantization}
\label{ssec:quantization}

Quantization is a technique for parameter reduction,
which reduces the model parameters to a lower precision (e.g., 8 and 16-bit integers)~\cite{shangguan2019optimizing}.
In quantization, we convert a floating point value $\mathrm{val_{float}}$ to an integer value $\mathrm{val_{quantized}}$ as follows:
\begin{align}
    \mathrm{val_{quantized}} = \mathrm{val_{float}} * s + z,
\end{align}
where $s$ and $z$ are the scale and offset parameters in Affine Quantization~\cite{wu2020integer}.
In this work, we apply dynamic quantization to the converted model,
which estimates these two parameters on-the-fly.
Since the parameters are calculated for each forward pass, it can speed up inference while preventing significant accuracy loss.
Quantization also contributes to the compression of the model size and the suppression of memory consumption while loading, as shown in Table~\ref{table:design:memory_consumption}.
We further investigate the effect of quantization on inference speed and memory usage, which is discussed more detail in Section~\ref{sec:experiment}.

\begin{table}
  \caption{
    Influence of quantization on model size in storage and memory consumption while loading. Conformer-based model with 83.2M parameters was used.
  }
  \centering
  \begin{tabular}{lcc}
    \toprule
    & Model size [Mb] & Memory consumption [Mb]\\
    \midrule
    Conformer & 378.4 & 750.8 \\
    w/ quantization & 165.4 & 338.8 \\
    \bottomrule
  \end{tabular}
\label{table:design:memory_consumption}
\end{table}

\section{Related Works}
\label{sec:related_works}

Much research has been conducted in speech processing to compress
the model size and improve inference speed.
The quantization technique has been actively studied throughout various tasks and architectures.
When applied to a trained ASR model,
quantization has little effect on Word Error Rate (WER) in the ASR task~\cite{shangguan2019optimizing,bie2019simplified,prasad2020quantization,kim2022integer}.
Xu et al.~\cite{xu2021mixed} has applied uniform and mixed precision quantization to neural language models used for ASR decoding.
They have achieved significant model compression with no degradation in accuracy.
More recently, Ding et al.~\cite{ding20224} has quantized Conformer-based ASR models and studied the influence on accuracy and inference speed.
Avila et al.~\cite{avila2022low} has shown the effectiveness of quantization on SLU models,
reporting that accuracy is not much affected.
In our work, we adopt the unified quantization method for various speech processing models,
including ASR, TTS, SLU, and ST,
and evaluate the effectiveness.

A few studies have focused on optimizing a DNN model at the application level.
Openja et al.~\cite{openja2022empirical} conducted a detailed investigation on ONNX conversion.
They have investigated accuracy degradation, inference time, and adversarial robustness for models before and after conversion to ONNX and CoreML library.
They have evaluated five image classification models, including two RNN-based and three CNN-based models, and concluded that all models have almost the same performance in ONNX model.
However, this paper does not mention about speech processing models, and to the best of our knowledge, there is no study that have conducted a detailed investigation in various speech processing tasks of ONNX format.
In this paper, we perform conversion with application-level optimizations for various speech processing tasks
to confirm the effectiveness of ONNX.
We will also confirm the effectiveness of ESPnet's model, which is highly accurate even under real-world conditions, for more practical research.

\section{Experiment}
\label{sec:experiment}

In this section, we compare PyTorch and ONNX models by evaluating their Real Time Factor (RTF) on various speech processing tasks (i.e., ASR, TTS, ST, and SLU).

\subsection{Automatic Speech Recognition}
\label{ssec:experiment_asr}

We used the test-clean dataset from the LibriSpeech corpus~\cite{panayotov2015librispeech} for measurement.
We compared the RTF and WER for the PyTorch and ONNX models.
We used the c6i.2xlarge instance for CPU inference and g5.xlarge instance for GPU from Amazon Elastic Compute Cloud.
For ONNX, we applied node fusion to the converted ONNX model to evaluate the effect.
Both PyTorch and fused models are quantized and evaluated as well.
We conducted evaluation with the following four major ASR models:
\begin{itemize}
    \item Transformer model (Trf-CTC/Att)
    \item Conformer model (Cfm-CTC/Att)
    \item Conformer model with HuBERT for feature extraction (Cfm-HuBERT)
    \item RNN Transducer model with ConformerEncoder (Cfm-Transducer)
\end{itemize}

\begin{table}[t]\centering
\caption{The comparison of mean and standard deviation (in brackets) of RTF and WER between PyTorch and ONNX model on CPU.}\label{table:asr:comparison_cpu}
\scriptsize
\begin{tabular}{lcccccc}\toprule
Model &ONNX & Fused & Quantized &RTF &WER [\%] \\
\midrule
\multirow{5}{*}{Trf-CTC/Att} & & & &0.379 (0.167) &3.4 \\
&\checkmark & & &0.226 (0.104) &3.5 \\
&\checkmark &\checkmark & &0.197 (0.071) &3.5 \\
& & &\checkmark &0.325 (0.121) &3.5 \\
&\checkmark &\checkmark &\checkmark &0.122 (0.040) &3.5 \\
\midrule
\multirow{5}{*}{Cfm-CTC/Att} & & & &0.417 (0.121) &2.3 \\
&\checkmark & & &0.426 (0.139) &2.1 \\
&\checkmark &\checkmark & &0.383 (0.103) &2.5 \\
& & &\checkmark &0.293 (0.104) &2.4 \\
&\checkmark &\checkmark &\checkmark &0.203 (0.052) &2.7 \\
\midrule
\multirow{5}{*}{Cfm-HuBERT} & & & &0.291 (0.055) &2.1 \\
&\checkmark & & &0.268 (0.059) &2.1 \\
&\checkmark &\checkmark & &0.260 (0.083) &2.1 \\
& & &\checkmark &0.245 (0.051) &2.0 \\
&\checkmark &\checkmark &\checkmark &0.149 (0.031) &2.1 \\
\midrule
\multirow{5}{*}{Cfm-Transducer} & & & &0.139 (0.006) &2.9 \\
&\checkmark & & &0.180 (0.249) &2.9 \\
&\checkmark &\checkmark & &0.179 (0.266) &2.9 \\
& & &\checkmark &0.135 (0.004) &3.0 \\
&\checkmark &\checkmark &\checkmark &0.116 (0.123) &3.2 \\
\bottomrule
\end{tabular}
\end{table}

\subsubsection{CPU Inference}
The results are shown in Table \ref{table:asr:comparison_cpu}.
We observed a significant speed up in the Transformer models.
In particular, the ONNX model with node fusion and quantization got $3.1 \times$ speed up from the original PyTorch model.
Conformer models also got $1.8 \times$ to $2.0 \times$ improvement from PyTorch model.

\subsubsection{Word Error Rate}
Table~\ref{table:asr:comparison_cpu} also shows that we did not get significant loss in WER.
Although there could be a little degradation between the vanilla and quantized models on both PyTorch and ONNX models,
there are no accuracy loss caused by the node fusion.

\begin{table}[t]\centering
\caption{
Real Time Factor on GPU.
}
\label{table:asr:comparison_gpu}
\scriptsize
\begin{tabular}{lcccc}\toprule
Model &ONNX & Fused  &RTF \\
\midrule
\multirow{3}{*}{Trf-CTC/Att} & & & 0.161 (0.041) \\
&\checkmark & & 0.091 (0.029) \\
&\checkmark &\checkmark & 0.086 (0.030) \\
\midrule
\multirow{3}{*}{Cfm-CTC/Att} & & & 0.175 (0.050) \\
&\checkmark & & 0.162 (0.060) \\
&\checkmark &\checkmark & 0.159 (0.061)
 \\
\midrule
\multirow{3}{*}{Cfm-HuBERT} & & & 0.129 (0.022) \\
&\checkmark & & 0.114 (0.051) \\
&\checkmark &\checkmark & 0.109 (0.051) \\
\midrule
\multirow{3}{*}{Cfm-Transducer} & & &0.188 (0.007) \\
&\checkmark & &0.168 (0.061) \\
&\checkmark &\checkmark & 0.166 (0.057) \\
\bottomrule
\end{tabular}
\end{table}

\subsubsection{GPU Inference}
We further investigated the conversion influence on GPU in Table~\ref{table:asr:comparison_gpu}.
We found that simply converting to ONNX alone did not result in a great speed up.
However, we observed a significant improvement with node fusion.
Node fusion achieves almost $1.6 \times$ faster in offline Transformer model.
Since ONNX Runtime does not support GPU inference with a quantized model with INT8 precision\footnote{The version of ONNX Runtime is v1.11.1.}, 
we did not evaluate the performance of quantization for GPU.

\subsubsection{Memory Usage}
We also evaluated the memory usage during inference on CPU.
The result is shown on Table~\ref{table:asr:memory}.
We found that the maximum memory usage after conversion was reduced 72\% in encoding process.
The memory consumption in the decoding process became 73\% lower. 
We noticed that the node fusion caused a little increase in memory usage.
Since several multiplications of tensors was fused into a single node,
the amount of memory allocated at once would increase.
Although the increase in memory usage can be seen in both encoding and decoding process,
the maximum memory consumption of vanilla and fused model was approximately 70\% lower compared to the PyTorch model.
We also found that quantization did not reduce memory consumption during inference.
In the quantized nodes, such as QAttention node, ONNX first set the bias to the output memory.
Since quantization was applied to multiplication of matrices in ONNX,
the bias was not quantized, which the Add operation was applied.
Therefore, the amount of memory that have been used during the execution was unchanged before and after quantization.

\begin{table}
  \caption{
    Maximum memory usage between PyTorch and ONNX models. We used Conformer based model without HuBERT.
  }
  \centering
  \begin{tabular}{llcc}
    \toprule
      \multicolumn{2}{l}{Library} &
      \multicolumn{1}{c}{Encode [Mb]} &
      \multicolumn{1}{c}{Decode [Mb]} \\
      \midrule
    
    PyTorch  & & 228.8 & 204.9 \\
    \midrule
    \multirow{3}{*}{ONNX} & Original & 36.4 & 56.1 \\
        & Fused & 40.6 & 60.3 \\
        & Quantized & 40.4 & 60.3 \\
    \bottomrule
  \end{tabular}
\label{table:asr:memory}
\end{table}

\subsection{Text-to-Speech}
\label{ssec:experiment_tts}

In the TTS task, we used 250 utterances from the LJSpeech dataset~\cite{ljspeech17} to measure the effectiveness of ONNX conversion.
We calculated RTF, Mel-Cepstral Distortion (MCD), and ASR-based Character Error Rate (CER) on the following four TTS models:
\begin{itemize}
    \item VITS~\cite{kim2021conditional}
    \item JETS~\cite{lim2022jets}
    \item Conformer-based FastSpeech2~\cite{ren2020fastspeech} with HiFiGAN~\cite{kong2020hifi}
    \item Conformer-based FastSpeech2 with MelGAN~\cite{kumar2019melgan}
\end{itemize}
Note that node fusion was not applied to the TTS models since they basically consist of CNN layers, which is not supported by node fusion logic.
We used four threads of CPUs (AMD EPYC 74F3, 3.2 GHz) for this evaluation.


The evaluation result is shown in Table~\ref{table:tts:compare}.
The result demonstrated that converting to the ONNX models was effective, resulting in approximately $1.3 \times$ speed up from the PyTorch models.
From CFS2 + HiFiGAN and CFS2 + MelGAN results, we can confirm that the calculation of neural vocoders, which mainly consists of CNN layers, is dominant in the inference of TTS models. 
Since VITS and JETS also have a similar component to HiFiGAN, the tendency should be common among all the models. 
We could not see the improvement by the quantization since we did not apply the quantization for CNN layers, which constitute most of the calculation time.
Focusing on the MCD and CER, we can confirm that the performance degradation caused by the conversion is limited, enabling us to keep the quality with faster inference.


\begin{table}\centering
\caption{The comparison of mean and standard deviation (in brackets) of RTF, MCD, and CER between PyTorch and ONNX model on CPU.}
\label{table:tts:compare}
\scriptsize
\begingroup
\scalefont{0.9}
\begin{tabular}{p{6em}ccccc}\toprule
Model &ONNX &Quantized &RTF &MCD [dB] &CER [\%] \\\midrule
\multirow{3}{*}{VITS} & & &0.273 (0.014) &6.852 (0.577) &2.1 \\
&\checkmark & &0.197 (0.013) &6.860 (0.566) &2.1 \\
&\checkmark &\checkmark &0.196 (0.011) &6.870 (0.575) &1.9 \\
\midrule
\multirow{3}{*}{JETS} & & &0.245 (0.012) &6.682 (0.537) &1.3 \\
&\checkmark & &0.183 (0.012) &6.682 (0.537) &1.3 \\
&\checkmark &\checkmark &0.183 (0.011) &6.684 (0.545) &1.3 \\
\midrule
\multirow{3}{*}{CFS2 + HiFiGAN} & & &0.264 (0.012) &6.468 (0.581) &1.2 \\
&\checkmark & &0.200 (0.008) &6.548 (0.573) &1.3 \\
&\checkmark &\checkmark &0.211 (0.007) &6.537 (0.581) &1.2 \\
\cdashlinelr{1-6}
\multirow{3}{*}{\ \ CFS2 only} & & &0.035 (0.004) &-- &-- \\
&\checkmark & &0.030 (0.003) &-- &-- \\
&\checkmark &\checkmark &0.032 (0.005) &-- &-- \\
\cdashlinelr{1-6}
\multirow{3}{*}{\ \ HiFiGAN only} & & &0.229 (0.010) &-- &-- \\
&\checkmark & &0.170 (0.008) &-- &-- \\
&\checkmark &\checkmark &0.179 (0.005) &-- &-- \\
\midrule
\multirow{3}{*}{CFS2 + MelGAN} & & &0.113 (0.013) &6.780 (0.560) &1.3 \\
&\checkmark & &0.111 (0.009) &6.843 (0.555) &1.3 \\
&\checkmark &\checkmark &0.111 (0.008) &6.807 (0.545) &1.2 \\
\cdashlinelr{1-6}
\multirow{3}{*}{\ \ CFS2 only} & & &0.033 (0.004) &-- &-- \\
&\checkmark & &0.031 (0.003) &-- &-- \\
&\checkmark &\checkmark &0.033 (0.005) &-- &-- \\
\cdashlinelr{1-6}
\multirow{3}{*}{\ \ MelGAN only} & & &0.080 (0.014) &-- &-- \\
&\checkmark & &0.081 (0.008) &-- &-- \\
&\checkmark &\checkmark &0.079 (0.005) &-- &-- \\
\bottomrule
\end{tabular}
\endgroup
\end{table}

\subsection{Speech Translation}
\label{ssec:experiment_st}

In the ST task, we used the IWSLT22 dialect dataset~\cite{anastasopoulos-etal-2022-findings},
which consists of Tunisian Arabic audio files and its transcriptions.
This is published for the Dialectual Speech Translation task, a shared task for IWSLT 2022.
The task aims to develop speech translation system for low-resource languages.
We report RTF and BiLingual Evaluation Understudy (BLEU) score with the Conformer model~\cite{yan-etal-2022-cmus}.
The evaluated model was built with ESPnet-ST~\cite{inaguma2020espnet}.
We used the c6i.2xlarge instance for CPU inference from Amazon Elastic Compute Cloud.

\begin{table}[t]\centering
\caption{The comparison of mean and standard deviation (in brackets) of RTF and BLEU score between PyTorch and ONNX model on CPU.}
\label{table:st:comparison}
\scriptsize
\begin{tabular}{cccccc}\toprule
ONNX & Fused  & Quantized  &RTF &BLEU \\
\midrule
& & & 0.134 (0.029) & 20.6 \\
\checkmark & & & 0.136 (0.038) & 20.5 \\
\checkmark &\checkmark & & 0.102 (0.022) & 20.5 \\
& &\checkmark &0.153 (0.031) & 20.8 \\
\checkmark &\checkmark &\checkmark & 0.075 (0.016) & 20.8 \\
\bottomrule
\end{tabular}
\end{table}

Table~\ref{table:st:comparison} shows that the effect of the conversion was also confirmed on the ST model,
with node fusion and quantization achieved approximately $1.9 \times$ speed up.
We observed that conversion does not have a significant effect on the BLEU score. 
Since the model architecture during inference was basically the same with ASR, 
we got similar performance improvement.


\subsection{Spoken Language Understanding}
\label{ssec:experiment_slu}

For the SLU task, we used the SLURP dataset~\cite{bastianelli2020slurp} for evaluation.
It consists of many prompts for an in-home personal robot assistant.
We randomly selected 200 audio samples from SLURP-real dataset.
We tested with the intent classification task and reported the RTF and accuracy as metrics.
The model is built with ESPnet-SLU~\cite{arora2022espnet} and consists of Conformer layers
We used the c6i.2xlarge instance for CPU inference from Amazon Elastic Compute Cloud.

Table~\ref{table:slu:comparison} shows the RTF and accuracy for intent classification task, and
we also observed an improvement in inference speed.
As with the ST models, the SLU models are basically the same architecture as in ASR.
Therefore, even for models other than Conformer,
similar speedups can be expected, as confirmed in section~\ref{ssec:experiment_asr}.
In the intent classification task, we could fasten inference without causing the degradation in accuracy.
The inference speed becomes $1.9 \times$ faster in quantized model
with no performance degradation.

\begin{table}[t]\centering
\caption{RTF and accuracy between PyTorch and ONNX model on CPU with 200 samples.}
\label{table:slu:comparison}
\scriptsize
\begin{tabular}{cccccc}\toprule
ONNX & Fused & Quantized  &RTF &Accuracy \\
\midrule
& & &0.215 (0.053) &0.909 \\
\checkmark & & &0.234 (0.083) &0.895 \\
\checkmark &\checkmark & &0.206 (0.070) &0.895 \\
& &\checkmark &0.199 (0.049) &0.915 \\
\checkmark &\checkmark &\checkmark &0.114 (0.035) &0.905 \\
\bottomrule
\end{tabular}
\end{table}

\subsection{Self-supervised Learning}
\label{ssec:experiment_ssl}

We also investigated applying our optimization techniques to speed up model training,
focusing on the usage of a pre-trained self-supervised acoustic model (PAM).
PAMs are often employed as feature extractors for training speech processing models,
where the pre-trained parameters can be used without fine-tuning~\cite{yang2021superb}.
We expect that converting a PAM to ONNX speeds up the feature extraction process and
makes the model training more efficient.
To that end, we compared PyTorch and ONNX versions of HuBERT~\cite{hsu2021hubert},
which was pre-trained on LibriLight-60k and applied to LibriSpeech-100h training.
We used a single V100 GPU with 16GB memory for measuring the training speed.
We observed a slight but clear speedup in the model training;
for one epoch of training, the PyTorch version took 1 hour and 53 minutes and the ONNX version took 1 hour 48 minutes.
The gain was smaller than those observed in the previous inference experiments
because the ONNX model seemed to be less effective when a large batchsize was used.

\section{Conclusion and Future work}
\label{sec:conclusion}

In this paper, we described a procedure for converting and optimizing complicated PyTorch-based models into ONNX format,
using the widely used speech processing toolkit, ESPnet, as an example.
By applying the conversion and optimization, including node fusion and quantization,
we achieved about $2 \times$ speedup in ASR, ST, and SLU tasks and $1.3 \times$ speedup in TTS task compared to the original PyTorch model.
We further investigated the effectiveness of fine-tuning process of a pre-trained acoustic model,
which revealed that the conversion does have effects on fine-tuning process.
In the future, we plan to support more tasks (e.g., Speech Enhancement, Speech Separation, and Voice Conversion)
and models (e.g., streaming).
Moreover, we plan to enhance the node fusion logic.

\section{Acknowledgement}
We thank the Extreme Science and Engineering Discovery Environment (XSEDE) \cite{xsede} supported by National Science Foundation grant number ACI-1548562. It uses the Bridges system \cite{nystrom2015bridges} supported by NSF award number ACI-1445606, at the Pittsburgh Supercomputing Center (PSC).



\begin{thebibliography}{10}

\bibitem{vaswani2017attention}
Ashish Vaswani, Noam Shazeer, Niki Parmar, Jakob Uszkoreit, Llion Jones,
  Aidan~N Gomez, {\L}ukasz Kaiser, and Illia Polosukhin,
\newblock ``Attention is all you need,''
\newblock {\em Proc. NeurIPS}, vol. 30, 2017.

\bibitem{gulati2020conformer}
Anmol Gulati, James Qin, Chung-Cheng Chiu, Niki Parmar, Yu~Zhang, Jiahui Yu,
  Wei Han, Shibo Wang, Zhengdong Zhang, Yonghui Wu, and Ruoming Pang,
\newblock ``Conformer: {C}onvolution-augmented {Transformer} for speech
  recognition,''
\newblock in {\em Proc. Interspeech}, 2020, pp. 5036--5040.

\bibitem{dong2018speech}
Linhao Dong, Shuang Xu, and Bo~Xu,
\newblock ``Speech-{Transformer}: {A} no-recurrence sequence-to-sequence model
  for speech recognition,''
\newblock in {\em Proc. ICASSP}, 2018, pp. 5884--5888.

\bibitem{karita2019comparative}
Shigeki Karita, Nanxin Chen, Tomoki Hayashi, Takaaki Hori, Hirofumi Inaguma,
  Ziyan Jiang, Masao Someki, Nelson Enrique~Yalta Soplin, Ryuichi Yamamoto,
  Xiaofei Wang, et~al.,
\newblock ``A comparative study on {Transformer} vs {RNN} in speech
  applications,''
\newblock in {\em Proc. ASRU}, 2019, pp. 449--456.

\bibitem{radfar2020end}
Martin Radfar, Athanasios Mouchtaris, and Siegfried Kunzmann,
\newblock ``End-to-end neural transformer based spoken language
  understanding,''
\newblock in {\em Proc. Interspeech}, 2020, pp. 866--870.

\bibitem{guo2021recent}
Pengcheng Guo, Florian Boyer, Xuankai Chang, Tomoki Hayashi, Yosuke Higuchi,
  Hirofumi Inaguma, Naoyuki Kamo, Chenda Li, Daniel Garcia-Romero, Jiatong Shi,
  et~al.,
\newblock ``Recent developments on {ESPnet} toolkit boosted by {Conformer},''
\newblock in {\em Proc. ICASSP}, 2021, pp. 5874--5878.

\bibitem{baevski2019vq}
Alexei Baevski, Steffen Schneider, and Michael Auli,
\newblock ``vq-wav2vec: Self-supervised learning of discrete speech
  representations,''
\newblock {\em arXiv preprint arXiv:1910.05453}, 2019.

\bibitem{baevski2020wav2vec}
Alexei Baevski, Yuhao Zhou, Abdelrahman Mohamed, and Michael Auli,
\newblock ``wav2vec 2.0: A framework for self-supervised learning of speech
  representations,''
\newblock {\em Proc. NeurIPS}, vol. 33, pp. 12449--12460, 2020.

\bibitem{hsu2021hubert}
Wei-Ning Hsu, Yao-Hung~Hubert Tsai, Benjamin Bolte, Ruslan Salakhutdinov, and
  Abdelrahman Mohamed,
\newblock ``Hubert: How much can a bad teacher benefit {ASR} pre-training?,''
\newblock in {\em Proc. ICASSP}, 2021, pp. 6533--6537.

\bibitem{yang2021superb}
Shu wen Yang, Po-Han Chi, Yung-Sung Chuang, Cheng-I~Jeff Lai, Kushal Lakhotia,
  Yist~Y. Lin, Andy~T. Liu, Jiatong Shi, Xuankai Chang, Guan-Ting Lin,
  Tzu-Hsien Huang, Wei-Cheng Tseng, Ko~tik Lee, Da-Rong Liu, Zili Huang, Shuyan
  Dong, Shang-Wen Li, Shinji Watanabe, Abdelrahman Mohamed, and Hung yi~Lee,
\newblock ``{SUPERB}: Speech processing universal {PERformance} benchmark,''
\newblock in {\em Proc. Interspeech}, 2021, pp. 1194--1198.

\bibitem{kuchaiev2019nemo}
Oleksii Kuchaiev, Jason Li, Huyen Nguyen, Oleksii Hrinchuk, Ryan Leary, Boris
  Ginsburg, Samuel Kriman, Stanislav Beliaev, Vitaly Lavrukhin, Jack Cook,
  et~al.,
\newblock ``{NeMo}: a toolkit for building ai applications using neural
  modules,''
\newblock {\em arXiv preprint arXiv:1909.09577}, 2019.

\bibitem{yao2021wenet}
Zhuoyuan Yao, Di~Wu, Xiong Wang, Binbin Zhang, Fan Yu, Chao Yang, Zhendong
  Peng, Xiaoyu Chen, Lei Xie, and Xin Lei,
\newblock ``{WeNet}: Production oriented streaming and non-streaming end-to-end
  speech recognition toolkit,''
\newblock {\em arXiv preprint arXiv:2102.01547}, 2021.

\bibitem{hanson1988comparing}
Stephen Hanson and Lorien Pratt,
\newblock ``Comparing biases for minimal network construction with
  back-propagation,''
\newblock {\em Proc. NeurIPS}, vol. 1, 1988.

\bibitem{lecun1989optimal}
Yann LeCun, John Denker, and Sara Solla,
\newblock ``Optimal brain damage,''
\newblock {\em Proc. NeurIPS}, vol. 2, 1989.

\bibitem{hassibi1992second}
Babak Hassibi and David Stork,
\newblock ``Second order derivatives for network pruning: Optimal brain
  surgeon,''
\newblock {\em Proc. NeurIPS}, vol. 5, 1992.

\bibitem{narang2017exploring}
Sharan Narang, Erich Elsen, Gregory Diamos, and Shubho Sengupta,
\newblock ``Exploring sparsity in recurrent neural networks,''
\newblock {\em arXiv preprint arXiv:1704.05119}, 2017.

\bibitem{lee2021layer}
Jaesong Lee, Jingu Kang, and Shinji Watanabe,
\newblock ``Layer pruning on demand with intermediate {CTC},''
\newblock {\em arXiv preprint arXiv:2106.09216}, 2021.

\bibitem{voita2019analyzing}
Elena Voita, David Talbot, Fedor Moiseev, Rico Sennrich, and Ivan Titov,
\newblock ``Analyzing multi-head self-attention: Specialized heads do the heavy
  lifting, the rest can be pruned,''
\newblock {\em arXiv preprint arXiv:1905.09418}, 2019.

\bibitem{michel2019sixteen}
Paul Michel, Omer Levy, and Graham Neubig,
\newblock ``Are sixteen heads really better than one?,''
\newblock {\em Proc. NeurIPS}, vol. 32, 2019.

\bibitem{li2019improving}
Sheng Li, Raj Dabre, Xugang Lu, Peng Shen, Tatsuya Kawahara, and Hisashi Kawai,
\newblock ``Improving {Transformer}-based speech recognition systems with
  compressed structure and speech attributes augmentation.,''
\newblock in {\em Proc. Interspeech}, 2019, pp. 4400--4404.

\bibitem{pang2018compression}
Ruoming Pang, Tara Sainath, Rohit Prabhavalkar, Suyog Gupta, Yonghui Wu,
  Shuyuan Zhang, and Chung-Cheng Chiu,
\newblock ``Compression of end-to-end models,''
\newblock in {\em Proc. Interspeech}, 2018, pp. 27--31.

\bibitem{takashima2018investigation}
Ryoichi Takashima, Sheng Li, and Hisashi Kawai,
\newblock ``An investigation of a knowledge distillation method for {CTC}
  acoustic models,''
\newblock in {\em Proc. ICASSP}, 2018, pp. 5809--5813.

\bibitem{graves2014towards}
Alex Graves and Navdeep Jaitly,
\newblock ``Towards end-to-end speech recognition with recurrent neural
  networks,''
\newblock in {\em Proc. ICML}, 2014, pp. 1764--1772.

\bibitem{gu2017non}
Jiatao Gu, James Bradbury, Caiming Xiong, Victor~OK Li, and Richard Socher,
\newblock ``Non-autoregressive neural machine translation,''
\newblock in {\em Proc. ICLR}, 2018.

\bibitem{chen2020non}
Nanxin Chen, Shinji Watanabe, Jes{\'u}s Villalba, Piotr {\.Z}elasko, and Najim
  Dehak,
\newblock ``Non-autoregressive {Transformer} for speech recognition,''
\newblock {\em IEEE Signal Process. Lett.}, vol. 28, pp. 121--125, 2020.

\bibitem{chan2020imputer}
William Chan, Chitwan Saharia, Geoffrey Hinton, Mohammad Norouzi, and Navdeep
  Jaitly,
\newblock ``Imputer: Sequence modelling via imputation and dynamic
  programming,''
\newblock in {\em Proc. ICML}, 2020, pp. 1403--1413.

\bibitem{higuchi2020mask}
Yosuke Higuchi, Shinji Watanabe, Nanxin Chen, Tetsuji Ogawa, and Tetsunori
  Kobayashi,
\newblock ``{Mask {CTC}: Non-Autoregressive End-to-End {ASR} with {CTC} and
  Mask Predict},''
\newblock in {\em Proc. Interspeech}, 2020, pp. 3655--3659.

\bibitem{higuchi2021comparative}
Yosuke Higuchi, Nanxin Chen, Yuya Fujita, Hirofumi Inaguma, Tatsuya Komatsu,
  Jaesong Lee, Jumon Nozaki, Tianzi Wang, and Shinji Watanabe,
\newblock ``A comparative study on non-autoregressive modelings for
  speech-to-text generation,''
\newblock in {\em Proc. ASRU}, 2021, pp. 47--54.

\bibitem{kumar2019melgan}
Kundan Kumar, Rithesh Kumar, Thibault de~Boissiere, Lucas Gestin, Wei~Zhen
  Teoh, Jose Sotelo, Alexandre de~Br{\'e}bisson, Yoshua Bengio, and Aaron~C
  Courville,
\newblock ``Melgan: Generative adversarial networks for conditional waveform
  synthesis,''
\newblock {\em Proc. NeurIPS}, vol. 32, 2019.

\bibitem{ren2020fastspeech}
Yi~Ren, Chenxu Hu, Xu~Tan, Tao Qin, Sheng Zhao, Zhou Zhao, and Tie-Yan Liu,
\newblock ``Fastspeech 2: Fast and high-quality end-to-end text to speech,''
\newblock {\em arXiv preprint arXiv:2006.04558}, 2020.

\bibitem{kim2021conditional}
Jaehyeon Kim, Jungil Kong, and Juhee Son,
\newblock ``Conditional variational autoencoder with adversarial learning for
  end-to-end text-to-speech,''
\newblock in {\em Proc. ICML}, 2021, pp. 5530--5540.

\bibitem{inaguma2021orthros}
Hirofumi Inaguma, Yosuke Higuchi, Kevin Duh, Tatsuya Kawahara, and Shinji
  Watanabe,
\newblock ``Orthros: Non-autoregressive end-to-end speech translation with
  dual-decoder,''
\newblock in {\em Proc. ICASSP}, 2021, pp. 7503--7507.

\bibitem{omachi2022non}
Motoi Omachi, Yuya Fujita, Shinji Watanabe, and Tianzi Wang,
\newblock ``Non-autoregressive end-to-end automatic speech recognition
  incorporating downstream natural language processing,''
\newblock in {\em Proc. ICASSP}, 2022, pp. 6772--6776.

\bibitem{tensorrt}
``{NVIDIA} {TensorRT},'' \url{https://developer.nvidia.com/tensorrt}, 2021,
\newblock [Online; Accessed on July-26-2022].

\bibitem{bai2019}
Junjie Bai, Fang Lu, Ke~Zhang, et~al.,
\newblock ``{ONNX}: Open neural network exchange,''
  \url{https://github.com/onnx/onnx}, 2019.

\bibitem{watanabe2018espnet}
Shinji Watanabe, Takaaki Hori, Shigeki Karita, Tomoki Hayashi, Jiro Nishitoba,
  Yuya Unno, Nelson {Enrique Yalta Soplin}, Jahn Heymann, Matthew Wiesner,
  Nanxin Chen, Adithya Renduchintala, and Tsubasa Ochiai,
\newblock ``{ESPnet}: End-to-end speech processing toolkit,''
\newblock in {\em Proceedings of Interspeech}, 2018, pp. 2207--2211.

\bibitem{shen2018natural}
Jonathan Shen, Ruoming Pang, Ron~J Weiss, Mike Schuster, Navdeep Jaitly,
  Zongheng Yang, Zhifeng Chen, Yu~Zhang, Yuxuan Wang, Rj~Skerrv-Ryan, et~al.,
\newblock ``Natural {TTS} synthesis by conditioning {WaveNet} on mel
  spectrogram predictions,''
\newblock in {\em Proc. ICASSP}, 2018, pp. 4779--4783.

\bibitem{onnxruntime}
ONNX~Runtime developers,
\newblock ``{ONNX} runtime,'' \url{https://onnxruntime.ai/}, 2021,
\newblock Version: x.y.z.

\bibitem{shaw-etal-2018-self}
Peter Shaw, Jakob Uszkoreit, and Ashish Vaswani,
\newblock ``Self-attention with relative position representations,''
\newblock in {\em Proc. NAACL HLT}, 2018, pp. 464--468.

\bibitem{dai-etal-2019-transformer}
Zihang Dai, Zhilin Yang, Yiming Yang, Jaime Carbonell, Quoc Le, and Ruslan
  Salakhutdinov,
\newblock ``Transformer-{XL}: Attentive language models beyond a fixed-length
  context,''
\newblock in {\em Proc. ACL}, 2019, pp. 2978--2988.

\bibitem{shangguan2019optimizing}
Yuan Shangguan, Jian Li, Qiao Liang, Raziel Alvarez, and Ian McGraw,
\newblock ``Optimizing speech recognition for the edge,''
\newblock {\em arXiv preprint arXiv:1909.12408}, 2019.

\bibitem{wu2020integer}
Hao Wu, Patrick Judd, Xiaojie Zhang, Mikhail Isaev, and Paulius Micikevicius,
\newblock ``Integer quantization for deep learning inference: Principles and
  empirical evaluation,''
\newblock {\em arXiv preprint arXiv:2004.09602}, 2020.

\bibitem{bie2019simplified}
Alex Bie, Bharat Venkitesh, Joao Monteiro, Md~Haidar, Mehdi Rezagholizadeh,
  et~al.,
\newblock ``A simplified fully quantized transformer for end-to-end speech
  recognition,''
\newblock {\em arXiv preprint arXiv:1911.03604}, 2019.

\bibitem{prasad2020quantization}
Amrutha Prasad, Petr Motlicek, and Srikanth Madikeri,
\newblock ``Quantization of acoustic model parameters in automatic speech
  recognition framework,''
\newblock {\em arXiv preprint arXiv:2006.09054}, 2020.

\bibitem{kim2022integer}
Sehoon Kim, Amir Gholami, Zhewei Yao, Nicholas Lee, Patrick Wang, Aniruddha
  Nrusimha, Bohan Zhai, Tianren Gao, Michael~W Mahoney, and Kurt Keutzer,
\newblock ``Integer-only zero-shot quantization for efficient speech
  recognition,''
\newblock in {\em Proc. ICASSP}, 2022, pp. 4288--4292.

\bibitem{xu2021mixed}
Junhao Xu, Jianwei Yu, Shoukang Hu, Xunying Liu, and Helen Meng,
\newblock ``Mixed precision low-bit quantization of neural network language
  models for speech recognition,''
\newblock {\em IEEE/ACM Trans. Audio, Speech, Lang. Process.}, vol. 29, pp.
  3679--3693, 2021.

\bibitem{ding20224}
Shaojin Ding, Phoenix Meadowlark, Yanzhang He, Lukasz Lew, Shivani Agrawal, and
  Oleg Rybakov,
\newblock ``4-bit {Conformer} with native quantization aware training for
  speech recognition,''
\newblock {\em arXiv preprint arXiv:2203.15952}, 2022.

\bibitem{avila2022low}
Anderson~R Avila, Khalil Bibi, Rui~Heng Yang, Xinlin Li, Chao Xing, and Xiao
  Chen,
\newblock ``Low-bit shift network for end-to-end spoken language
  understanding,''
\newblock {\em arXiv preprint arXiv:2207.07497}, 2022.

\bibitem{openja2022empirical}
Moses Openja, Amin Nikanjam, Ahmed~Haj Yahmed, Foutse Khomh, Zhen Ming, et~al.,
\newblock ``An empirical study of challenges in converting deep learning
  models,''
\newblock {\em arXiv preprint arXiv:2206.14322}, 2022.

\bibitem{panayotov2015librispeech}
Vassil Panayotov, Guoguo Chen, Daniel Povey, and Sanjeev Khudanpur,
\newblock ``Librispeech: An {ASR} corpus based on public domain audio books,''
\newblock in {\em Proc. ICASSP}, 2015, pp. 5206--5210.

\bibitem{ljspeech17}
Keith Ito and Linda Johnson,
\newblock ``The {LJ} speech dataset,''
  \url{https://keithito.com/LJ-Speech-Dataset/}, 2017.

\bibitem{lim2022jets}
Dan Lim, Sunghee Jung, and Eesung Kim,
\newblock ``{JETS}: Jointly training {FastSpeech2} and {HiFi-GAN} for end to
  end text to speech,''
\newblock {\em arXiv preprint arXiv:2203.16852}, 2022.

\bibitem{kong2020hifi}
Jungil Kong, Jaehyeon Kim, and Jaekyoung Bae,
\newblock ``{HiFi-GAN}: Generative adversarial networks for efficient and high
  fidelity speech synthesis,''
\newblock {\em Proc. NeurIPS}, vol. 33, pp. 17022--17033, 2020.

\bibitem{anastasopoulos-etal-2022-findings}
Antonios Anastasopoulos, Lo{\"\i}c Barrault, Luisa Bentivogli, Marcely
  Zanon~Boito, Ond{\v{r}}ej Bojar, Roldano Cattoni, Anna Currey, Georgiana
  Dinu, Kevin Duh, Maha Elbayad, Clara Emmanuel, Yannick Est{\`e}ve, Marcello
  Federico, Christian Federmann, Souhir Gahbiche, Hongyu Gong, Roman
  Grundkiewicz, Barry Haddow, Benjamin Hsu, D{\'a}vid Javorsk{\'y}, V{\u{e}}ra
  Kloudov{\'a}, Surafel Lakew, Xutai Ma, Prashant Mathur, Paul McNamee, Kenton
  Murray, Maria N{\v{a}}dejde, Satoshi Nakamura, Matteo Negri, Jan Niehues,
  Xing Niu, John Ortega, Juan Pino, Elizabeth Salesky, Jiatong Shi, Matthias
  Sperber, Sebastian St{\"u}ker, Katsuhito Sudoh, Marco Turchi, Yogesh Virkar,
  Alexander Waibel, Changhan Wang, and Shinji Watanabe,
\newblock ``Findings of the {IWSLT} 2022 evaluation campaign,''
\newblock in {\em Proceedings of the 19th International Conference on Spoken
  Language Translation (IWSLT 2022)}, Dublin, Ireland (in-person and online),
  May 2022, pp. 98--157, Association for Computational Linguistics.

\bibitem{yan-etal-2022-cmus}
Brian Yan, Patrick Fernandes, Siddharth Dalmia, Jiatong Shi, Yifan Peng, Dan
  Berrebbi, Xinyi Wang, Graham Neubig, and Shinji Watanabe,
\newblock ``{CMU}{'}s {IWSLT} 2022 dialect speech translation system,''
\newblock in {\em Proceedings of the 19th International Conference on Spoken
  Language Translation (IWSLT 2022)}, Dublin, Ireland (in-person and online),
  May 2022, pp. 298--307, Association for Computational Linguistics.

\bibitem{inaguma2020espnet}
Hirofumi Inaguma, Shun Kiyono, Kevin Duh, Shigeki Karita, Nelson Enrique~Yalta
  Soplin, Tomoki Hayashi, and Shinji Watanabe,
\newblock ``{ESPnet-ST}: All-in-one speech translation toolkit,''
\newblock {\em arXiv preprint arXiv:2004.10234}, 2020.

\bibitem{bastianelli2020slurp}
Emanuele Bastianelli, Andrea Vanzo, Pawel Swietojanski, and Verena Rieser,
\newblock ``{SLURP}: A spoken language understanding resource package,''
\newblock {\em arXiv preprint arXiv:2011.13205}, 2020.

\bibitem{arora2022espnet}
Siddhant Arora, Siddharth Dalmia, Pavel Denisov, Xuankai Chang, Yushi Ueda,
  Yifan Peng, Yuekai Zhang, Sujay Kumar, Karthik Ganesan, Brian Yan, et~al.,
\newblock ``{ESPnet-SLU}: Advancing spoken language understanding through
  {ESPnet},''
\newblock in {\em Proc. ICASSP}, 2022, pp. 7167--7171.

\bibitem{xsede}
J.~Towns, T.~Cockerill, M.~Dahan, I.~Foster, K.~Gaither, A.~Grimshaw,
  V.~Hazlewood, S.~Lathrop, D.~Lifka, G.~D. Peterson, R.~Roskies, J.~R. Scott,
  and N.~Wilkins-Diehr,
\newblock ``{XSEDE}: Accelerating scientific discovery,''
\newblock {\em Computing in Science \& Engineering}, vol. 16, no. 5, pp.
  62--74, 2014.

\bibitem{nystrom2015bridges}
Nicholas~A Nystrom, Michael~J Levine, Ralph~Z Roskies, and J~Ray Scott,
\newblock ``Bridges: a uniquely flexible {HPC} resource for new communities and
  data analytics,''
\newblock in {\em Proc. {XSEDE}}, 2015, pp. 1--8.

\end{thebibliography}
\end{document}